\documentclass[10pt]{iopart}

\usepackage{iopams,amsfonts,amssymb,amsthm} 

\usepackage{latexsym}
\usepackage{rawfonts}
\usepackage{multirow}
\usepackage{rotating}
\usepackage{lscape}
\usepackage{graphicx}\graphicspath{{figs/}}
\usepackage{subfigure}
\usepackage[usenames,dvipsnames]{color}

\input{prepictex}
\input{pictex}
\input{postpictex}

\definecolor{Maroon}{rgb}{0.70,0.0,0.0}
\definecolor{Maroon1}{rgb}{0.40,0.0,0.0}
\definecolor{Brown}{rgb}{0.7,0.3,0}
\definecolor{Navy}{rgb}{0.3,0.0,0.4}
\definecolor{Green}{cmyk}{1,0,1,0.2}
\definecolor{Red}{cmyk}{0,1,1,0}
\definecolor{DarkRed}{cmyk}{0,1,1,0.6}
\definecolor{DarkBlue}{cmyk}{1,1,0,0.5}
\definecolor{DarkGreen}{cmyk}{1,0,1,0.65}
\definecolor{OrangeRed}{cmyk}{0,1,0.87,0}
\definecolor{RedOrange}{cmyk}{0,0.77,0.87,0}
\definecolor{Orange}{cmyk}{0,0.61,0.87,0}
\definecolor{Offwhite}{cmyk}{.07,.15,.15,0}
\definecolor{Offwhite2}{cmyk}{.04,.02,.03,0}

\newtheorem{theorem}{{Theorem}}

\newtheorem{lemma}[theorem]{{Lemma}}

\newcommand{\q}{\quad}
\newcommand{\qq}{\qquad}

\newcommand{\Ref}[1]{(\ref{#1})}

\newcommand{\IntN}{{\mathbb{Z}}}

\def\L{\left(}
\def\R{\right)}
\def\LH{\left[}
\def\RH{\right]}
\def\LC{\left\{}
\def\RC{\right\}}

\def\LV{\left|}
\def\RV{\right|}
\def\lfl{\left\lfloor}
\def\rfl{\right\rfloor}
\def\lcl{\left\lceil}
\def\rcl{\right\rceil}
\def\vert{{\,\hbox{\large$|$}\,}}

\def\C#1{{\cal #1}}

\def\eps{\epsilon}

\def\Bi#1#2{ \L {{#1}\atop{#2}} \R }

\def\m#1{{m_#1}}

\def\sfrac#1#2{\hbox{\normalsize $\frac{#1}{#2}$}}

\def\half{{\sfrac{1}{2}}}

\begin{document}

\title[]{The pressure exerted by adsorbing directed lattice paths and staircase
polygons}

\author{
E.J. Janse van Rensburg$^1$\footnote[1]{\texttt{rensburg@yorku.ca}}
and T. Prellberg$^2$\footnote[2]{\texttt{t.prellberg@qmul.ac.uk}}
}

\address{$^1$Department of Mathematics and Statistics, 
York University, Toronto, Ontario M3J~1P3, Canada\\}

\address{$^2$School of Mathematical Sciences,
Queen Mary, University of London,
Mile End Road, London E1~4NS, United Kingdom}

\begin{abstract}
A directed path in the vicinity of a hard wall exerts pressure on the
wall because of loss of entropy.  The pressure at a particular point may 
be estimated by estimating the loss of entropy if the point is excluded 
from the path.  In this paper we determine asymptotic expressions for 
the pressure on the $X$-axis in models of adsorbing directed
paths in the first quadrant.  Our models show that the pressure 
vanishes in the limit of long paths in the desorbed phase, but there is a 
non-zero pressure in the adsorbed phase.  We determine asymptotic
approximations of the pressure for finite length Dyck paths and directed paths,
as well as for a model of adsorbing staircase polygons with both ends grafted
to the $X$-axis.
\end{abstract}

\submitto{\JPA}
\maketitle

\section{Introduction}

A linear polymer attached to a hard wall by an endpoint loses 
entropy due to the presence of the wall.  The loss of entropy
induces a repulsive force on the wall.  Such forces have 
been measured experimentally \cite{BHGMLSW95,CS95,CNC03} and 
decay with distance from the point where the polymer is attached. 

A simple model of the above is a directed path from the origin
in the first quadrant of the square lattice.  More precisely, let 
$d_n^+$ be the number of directed paths from the origin
given North-East and South-East steps in the half square lattice
$\IntN^2_+ =  \{ (n,m)\in\IntN^2 \vert \hbox{$m\geq 0$}\}$. 
Since the path loses entropy due to the boundary of $\IntN^2_+$
(the $X$-axis) there is a net pressure on the $X$-axis (which we
shall also call a \textit{hard wall}).

The entropy of the paths is given by $S_n^+ = k_B\log d_n^+$, where
$k_B$ is Boltzmann's constant. The pressure at a point $x = (N,0)$ 
in the hard wall may be computed by calculating the reduction in entropy
if paths passing though this point are excluded from the ensemble.  

Let the number of paths avoiding the point $x$ be denoted by 
$d_n^+(x)$. Then the loss of entropy is given by
\begin{equation}
\Delta S^+_n(x) = k_B\log d_n^+(x) - k_B\log d_n^+ .
\end{equation}
In this case the change in free energy of the model is also given
by $\Delta \C{F}_n (T) = -T\,\Delta S^+ (x)$ at temperature $T$.
Hence the pressure of the path on the wall can be estimated
by computing the change in extensive free energy if the point 
$x$ is excluded and thereby changing the volume by $\Delta V_x$.  

That is, the pressure on the point $x$ is given by the 
discrete derivative of $\C{F}_n(T)$ to the volume element 
containing $x$:
\begin{equation}
P_n(x) = -\sfrac{\Delta \C{F}_n (T)}{\Delta V_x}
= \sfrac{k_B T\,\Delta S^+_n(x)}{\Delta V_x}. 
\end{equation}
In two dimensions, $\Delta V_x$ will be an area element containing
the point $x$.

Adopt units $k_B T=1$ and $\Delta V_x=1$ to see that the pressure
at the point $x$ in this model is given by
\begin{equation}
P_n(x) = \log d_n^+(x) - \log d_n^+ .
\label{eqnW1}  
\end{equation}
In other words, the pressure at $x$ is the discrete derivative 
of the extensive free energy with respect to a unit change in 
volume at the point $x$ in reduced units.  

Observe that we use the convention that $P_n(x)$ is negative.

More generally, the paths may be interacting with the hard wall
by adsorbing in it, or may be interacting in some other way. 
In this case the above is a directed model of a polymer 
adsorbing in the hard wall, and $P_n(x)$ is the pressure due
to presence of the hard wall.

If $k_BT=1$ as above, then the interaction strength of the
path with the wall is given by an activity $z$, and the partition 
function $Z_n^+(z)$ of the model gives the extensive free energy 
$\C{F}_n(z) = \log Z_n^+(z)$.  The pressure at a point $x$
in the hard wall is then obtained as above, and is given by
\begin{equation}
P_n(x) = \log Z_n^+(z;x) - \log Z_n^+(z) 
\label{eqnW2}  
\end{equation}
where $Z_n^+(z;x)$ is the partition function of the ensemble of
paths which avoids the point $x$ is excluded from the ensemble.

The above approach was used for a square lattice self-avoiding walk 
model of linear polymers grafted to a hard wall  \cite{DJ12}. 
In this exact enumeration study the data show that pressure 
decreases with distance from the origin. 
 
In this paper our intension is to add to these self-avoiding
walk results by examining a directed version of this model. We
shall in particular look at the asymptotic behaviour of the 
pressure as function of the length of the directed walk and the 
location of pressure point $x$. We shall also extend our 
results to include adsorbing directed paths, as well as a model 
of an adsorbing staircase polygon.

\subsection{The pressure due to directed paths}

The number of directed paths of $n$ steps from the origin in
the square lattice $\IntN^2$ is $P_n = 2^n$.  If the path is 
restricted to the half lattice $\IntN^2_+$, then it is
a \textit{positive path}, and if it is further constrained
by having to end in the $X$-axis (which is the hard wall),
then it is a \textit{Dyck path} \cite{S99}.  Examples
are illustrated in figures \ref{figureA}(a) and (b).

The number of positive paths of length $n$ is given by
\begin{equation}
T_{2n} = \Bi{2n}{n}\q\hbox{and}\q T_{2n+1} = \Bi{2n+1}{n+1},
\end{equation}
and the number of Dyck paths of length $2n$ is given by
\begin{equation}
D_{2n} = \frac{1}{n+1}\Bi{2n}{n} .
\end{equation}
From these expressions one may determine that the number of 
Dyck paths passing through the point $x$ with coordinates 
$x=(2N,0)$.  This is given by
\begin{equation}
D_{2N}\cdot D_{2n-2N} = \frac{1}{(N+1)(n-N+1)}
\Bi{2N}{N} \Bi{2(n-N)}{n-N} .
\end{equation}
Thence, one may determine the pressure on the hard wall at
the point  $x$  from equation \Ref{eqnW1} to be
\begin{equation}
\fl
P_{2n}^D(2N) = \log\L D_{2n} - D_{2N}\cdot D_{2n-2N} \R
             - \log D_{2n}
= \log \L 1 - \sfrac{D_{2N}\cdot D_{2n-2N}}{D_{2n}} \R.
\end{equation}
Putting $N = \lfl a\,n\rfl$ for some $a\in (0,1)$ and expanding
asymptotically in $n$ gives
\begin{equation}
P_{2n}^D(2\lfl a\,n\rfl) = - \frac{1}{\sqrt{\pi n^3 a^3 (1-a)^3}} + 
O\L n^{-5/2}\R .
\label{eqnDD1}   
\end{equation}
Observe that $P_{2n}^D(2\lfl a\,n\rfl) = O\L n^{-3/2}\R$ and
that if pressure is rescaled by $n^{3/2}$, then one obtains
\begin{equation}
p^D(a) = \lim_{n\to\infty} \L n^{3/2} P_{2n}^D(2\lfl a\,n\rfl) \R
= - \frac{1}{\sqrt{\pi a^3 (1-a)^3}} ,
\label{eqnpDD1}   
\end{equation}
as a residual rescaled pressure in the scaling limit.

A similar calculation may be done with positive paths.
In this case the pressure on the point $(2N,0)$ is given by
\begin{equation}
\fl
P^T_{2n}(2N) = \log\L T_{2n} - D_{2N}\cdot T_{2n-2N} \R
             - \log T_{2n}
= \log \L 1 - \sfrac{D_{2N}\cdot T_{2n-2N}}{T_{2n}} \R.
\end{equation}
Expanding asymptotically, and keeping the leading order 
term gives
\begin{equation}
P^T_{2n}(2a\,n) = - \frac{1}{\sqrt{\pi n^3 a^3 (1-a)}} + 
O\L n^{-5/2}\R 
\label{eqnDD2}   
\end{equation}
for pressure of positive paths of length $2n$ at a distance
$2N = 2\lfl a\,n \rfl$ from the origin.  

Observe that $P_{2n}^T(2\lfl a\,n\rfl) = O\L n^{-3/2}\R$ (as
was seen for Dyck paths) and again rescaling the pressure 
by $n^{3/2}$ one obtains
\begin{equation}
p^T(a) = \lim_{n\to\infty} \L n^{3/2} P_{2n}^T(2\lfl a\,n\rfl) \R
= - \frac{1}{\sqrt{\pi a^3 (1-a)}} ,
\label{eqnpDD2}   
\end{equation}
as the residual rescaled pressure in the scaling limit.

We shall generalise the results in equations \Ref{eqnDD1} and
\Ref{eqnDD2} to models of adsorbing Dyck paths, adsorbing directed
paths, and a model of adsorbing staircase polygons grafted to the
$X$-axis.  While we extract the infinite $n$ behaviour in each model,
we will mostly be concerned with finite $n$ behaviour, which we 
shall approximate by asymptotic expressions.

\begin{figure}[t!]
\centering \hfil
\input{figureA.tex}
\caption{Two models of an adsorbing polymer grafted to a hard wall.  (a)  A
directed path from the origin in $\IntN^2_+$.
The path interacts with the $X$-axis with an activity $z>0$.
If $z>1$ then the path is attracted to the $X$-axis, if $z<1$
it is repelled.  The path exerts a force on vertices of the $X$-axis.
(b) This model is similar to (a), but both endpoints of the path
are constrained to lie in the $X$-axis.  This is a model of 
adsorbing Dyck paths.  Observe that we use the convention that
the vertex at the origin is not weighted by $z$. }
\label{figureA}  
\end{figure}

\subsection{Adsorbing paths and staircase polygons}

%

A directed path with steps $(1,-1)$ and $(1,1)$ from the origin 
in the half lattice $\IntN^2_+$ is illustrated in figure \ref{figureA}(a).
Vertices of the path in the \textit{adsorbing line} $Y=0$ are
called \textit{visits}, and they are weighted with the generating
variable $z$ (which is related to temperature by $z = e^{1/k_BT}$
in lattice units).  By convention, the origin, although itself a visit,
is not weighted.  A directed path with weighted visits is an 
\textit{adsorbing directed path}.

If the directed path is constrained to end in the adsorbing line,
then it is a Dyck path (see figure \ref{figureA}(b)). A Dyck path with
weighted visits is an \textit{adsorbing Dyck path}  \cite{BEO98,W98}.

The model in figure \ref{figureA}(b) is that of an adsorbing
Dyck path \cite{BEO98}.  If the partition function of this model
is denoted by $D_n(z)$, then the generating function of the model
may be evaluated:
\begin{equation}
g(t,z) = \sum_{n=0}^\infty D_n(z)\,t^n 
= \frac{2}{2-z(1-\sqrt{1-4t^2})} ,
\end{equation}
where the generating variable $t$ is conjugate the length $n$ (the
number of steps in the path).  The \textit{intensive}
limiting free energy of the model is defined by
\begin{equation}
\C{F}_D (z) = \lim_{n\to\infty} \sfrac{1}{n} \log D_n(z)
\end{equation}
and by comparison to $g(t,z)$ it follows that $\C{F}_D(z)
= - \log t_c(z)$, where $t_c(z)$ is the radius of convergence
of $g(t,z)$.  One may explicitly compute this from the above:
\begin{equation}
\C{F}_D(z) = \cases{
\log 2, &\hbox{if $z\leq 2$}; \\
\log z - \sfrac{1}{2} \log(z-1), &\hbox{if $z>2$}.}
\end{equation}
In other words, $\C{F}_D(z)$ is non-analytic at $z_c^+=2$.
For $z<2$ the density of visits (given by $\C{E}(z) = z\sfrac{d}{dz}
\C{F}_D(z)$) is zero -- this is the \textit{desorbed phase}.  
For $z>2$ this is $\C{E}(z) = \sfrac{z-2}{2(z-1)}
>0$ and the model is said to be in an \textit{adsorbed phase}
since the density of visits to the hard wall is positive.

Similar calculations can be done for the model in figure
\ref{figureA}(a), and the critical point is also at $z_c^+=2$
with a desorbed phase for $z<2$ and an adsorbed phase
for $z>2$.

The partition functions of adsorbing directed and Dyck paths
are known to be given by
\begin{equation}
T_n(z) = \sum_{m=0}^{\lfl n/2 \rfl} \Bi{n}{\lcl n/2\rcl+m}\,(z-1)^m
\label{eqn1} 
\end{equation}
for adsorbing directed paths, and
\begin{equation}
D_n(z) = \sum_{m=0}^{\lfl n/2 \rfl} \frac{4m+2}{n+2(m+1)}
\Bi{n}{\lfl n/2\rfl+m}\,(z-1)^m
\label{eqn2} 
\end{equation}
for Dyck paths, see for example references \cite{BEO98,JvR00}.

The \textit{extensive} free energies of these models are defined by
\begin{equation}
\C{F}^T_n  (z) = \log T_n(z),\q\hbox{and}\q
\C{F}^D_n (z) = \log D_n (z) .
\end{equation}

If $(q,0)$ is a vertex in the adsorbing line $Y=0$, then define,
similar to the above, the free energies of paths which avoids $(q,0)$,
and denote these by $\C{F}^T_n(z;q)$ and $\C{F}^D_n (z;q)$.

By equation \Ref{eqnW1} the pressure on the vertex $(q,0)$ is given 
by the free energy differences
\begin{equation}
\fl \hspace{1cm}
P_n^T(z;q) = \C{F}^T_n(z;q) -\C{F}^T_n(z)  \q\hbox{and}\q
P_n^D(z;q) = \C{F}^D_n(z;q) -\C{F}^D_n(z) 
\label{eqn4} 
\end{equation}
in each model.  Observe that $P_n^T(z;q) = P_n^D(z;q) = 0$ 
if $q$ is odd, so that a non-zero force can be obtained only for even values of $q$. 

For both models we shall show that for $a\in(0,1)$,
\begin{equation}
\fl
\lim_{n\to\infty} P_n^T(z;2\lfl an/2\rfl) =
\lim_{n\to\infty} P_n^D(z;2\lfl an/2\rfl) =
\cases{
0 ,& \hbox{if $z\leq 2$;} \\
-\log(z-1), & \hbox{if $z>2$.}
}
\end{equation}
For finite values of $n$, we determine asymptotic approximations for
$P_n^T(z;2\lfl an/2\rfl)$ and $P_n^D(z;2\lfl an/2\rfl)$ (for $n$ even). These are given by
\begin{equation}
\fl
P_n^T(z;2\lfl an/2 \rfl) \simeq
\cases{
-\frac{8}{\sqrt{2\pi n^3a^3(1-a)}\, \log^2(z-1)},
   & \hbox{if $z<2$}; \\
-\frac{\sqrt{2}}{\sqrt{\pi n a}} 
   & \hbox{if $z=2$}; \\
-\log\L z-1 \R
 - \frac{A_T}{12na(1-a)(z-1)(z-2)},
   & \hbox{if $z>2$}; \\
}
\label{eqn20AA}  
\end{equation}
where 
\[ A_T = (a^2-a+1)(z^4 + 8z^3 + 30z^2-32z+16) -6a^2z^2 , \]
for the model of directed paths.  In the case of adsorbing Dyck paths,
\begin{equation}
\fl
P_n^D(z;2\lfl an/2 \rfl) \simeq
\cases{
-\frac{8}{\sqrt{2\pi n^3a^3(1-a)^3}\, \log^2(z-1) },
   & \hbox{if $z<2$}; \\
-\frac{\sqrt{2}}{\sqrt{\pi n a(1-a)}}
   & \hbox{if $z=2$}; \\
-\log\L z-1 \R
 - \frac{A_D}{12na(1-a)(z-1)(z-2)},
   & \hbox{if $z>2$}; \\
}
\label{eqn21AA}  
\end{equation}
for even values of $n$, where 
\[ A_D = (a^2-a+1)(z^4 + 8z^3 + 30z^2-32z+16) .\]
From these results one can easily extract the residual rescaled
pressures similar to equations \Ref{eqnpDD1} and \Ref{eqnpDD2}.
However, notice the different scaling in $n$ in the different regimes:
In the desorbed phase the pressure is of order $O\L n^{-3/2} \R$,
at the critical adsorption point $O\L n^{-1/2} \R$ and in the
adsorbed phase $- \log (z-1) + O\L n^{-1} \R$.

In the case of a model of adsorbing staircase polygons with
both endpoints grafted to the adsorbing line (figure \ref{figureG}(a)),
a similar asymptotic analysis gives the following expressions for the 
pressure:
\begin{equation}
\fl
P_n^S(z;2\lfl an/2 \rfl) \simeq
\cases{
-\frac{5z}{2n^{3/2}(2-z)^2 \sqrt{\pi a^3(1-a)^3}},
   & \hspace{-8mm} \hbox{if $z<2$}; \\
-\frac{3}{2\sqrt{\pi n a(1-a)}} ,
   & \hspace{-8mm} \hbox{if $z=2$}; \\
-\log\L z-1 \R
 - \frac{3(4z^2-29z+16)}{8n(z-2)} \L 1+\sfrac{9z}{8n(z-2)^2}\R \\
 - \frac{3z^2}{2\sqrt{\pi n^3 a^3(1-a)^3}(z-1)^2(z-2)\LV \log(z-1) \RV},
   & \hspace{-8mm} \hbox{if $z>2$}. \\
}
\label{eqn22AA}  
\end{equation}
From these expressions one similarly obtains that there are
different scalings in the different regimes:
In the desorbed phase the pressure is of order $O\L n^{-3/2} \R$,
at the critical adsorption point $O\L n^{-1/2} \R$ and in the
adsorbed phase $- \log (z-1) + O\L n^{-1} \R$.

\section{The forces exerted by adsorbing directed paths in a half-space}

The partition function of adsorbing directed paths which
pass through the vertex $(q,0)$ is given by
$D_q(z)\,T_{n-q}(z)$ (by equations \Ref{eqn1} and \Ref{eqn2}).
Thus the partition function of paths avoiding the vertex $(q,0)$ is 
$T_n(q) - D_q(z)\,T_{n-q}(z)$.
Substitution into equation \Ref{eqn4} and simplifying gives
\begin{equation}
P_n^T(z;q) = \log \L 1 - \sfrac{D_q(z)\,T_{n-q}(z)}{T_n(z)} \R .
\label{eqn5}  
\end{equation}
Observe that $P_n^T(z;q)$ is negative, since the pressure is directed
onto the vertex $(q,0)$ from above (and in the negative direction).

A similar argument shows that
\begin{equation}
P_n^D(z;q) = \log \L 1 - \sfrac{D_q(z)\,D_{n-q}(z)}{D_n(z)} \R 
\label{eqn6} 
\end{equation}
in the case of adsorbing Dyck paths.

A result in reference \cite{BEO98}
shows that for $z>0$ the partition function $D_n (z)$ may 
be expressed in as 
\begin{equation}
D_{2n} (z) = \frac{z-2}{z-1} \L \frac{z^2}{z-1}\R^{n} \theta(z-2)
+ \frac{1}{z}\sum_{s={n}}^\infty C_s\, \L \frac{z-1}{z^2} \R^{s-n}
\label{eqn7a}  
\end{equation}
where the $C_s = \sfrac{1}{s+1} \Bi{2s}{s}$ are Catalan numbers
and $\theta$ is the Heaviside step function.
The summation can be bound as follows:  $C_s \leq 4^s$ and thus
\begin{equation}
\sum_{s=n}^\infty C_s\, \L \frac{z-1}{z^2} \R^{s-n} 
\leq 4^n \sum_{s=n}^\infty 4^{s-n}\, \L \frac{z-1}{z^2} \R^{s-n} 
= \frac{4^n z^2}{(z-2)^2} .
\end{equation}
This shows that for $z\geq 2$
\begin{equation}
\hspace{-2.5cm}
\frac{z-2}{z-1} \L \frac{z^2}{z-1}\R^{n} \leq D_{2n} (z) \leq
\frac{z-2}{z-1} \L \frac{z^2}{z-1}\R^{n}\L 1 + 
\frac{z(z-1)}{(z-2)^3} \L \frac{4(z-1)}{z^2} \R^n \R .
\label{eqn9b}  
\end{equation}
Since $4 < z^2/(z-1)$ if $z>2$, the above proves that
\begin{equation}
D_n (z) = \frac{z-2}{z-1} \L \frac{z}{\sqrt{z-1}}\R^n
(1+ o(1)) ,\q\hbox{if $z> 2$.}
\label{eqn10A}  
\end{equation}

The result in equation \Ref{eqn10A} can be substituted 
into equation \Ref{eqn6} to give the following lemma:
\begin{lemma}
The limiting force on the adsorbing line by a Dyck path if $z>2$ is 
\[\lim_{n\to\infty} P_n^D(z;2\lfl an/2\rfl)
= - \log (z-1). \]
for any $a\in(0,1)$, where the limit is taken through even values of $n$.
\qed
\label{lemma1}  
\end{lemma}

A similar argument may be made for directed paths, using the
following representation for the partition function
\begin{equation}
\hspace{-2cm}
T_{2n}(z) = \L \frac{z^2}{z-1} \R^n \theta(z-2)
+ \frac{1}{z^2} \sum_{s=n}^\infty C_s \,(1-s(z-2))\,
\L \frac{z-1}{z^2}\R^{s-n}
\end{equation}
from reference \cite{BEO98}.  This shows that
\begin{equation}
T_{n} (z) = 
\L \frac{z}{\sqrt{z-1}} \R^n (1+ o(1)), \q\hbox{if $z>2$} .
\label{eqn12B}  
\end{equation}
Substitution of equations \Ref{eqn10A} and \Ref{eqn12B}
into equation \Ref{eqn5} shows that
\begin{lemma}
The limiting force on the adsorbing line by a directed path if $z>2$ is 
\[\lim_{n\to\infty}  P_n^T(z;2\lfl an/2\rfl)
= - \log (z-1). \]
for any $a\in(0,1)$.
\qed
\label{lemma2}  
\end{lemma}

Lemmas \ref{lemma1} and \ref{lemma2} show that the limiting
force in the regime $z>2$ is $-\log (z-1)$ in both models.
Observe that this force approaches zero as $z\to 2^+$.  By equation
\Ref{eqn4} the force cannot be positive, and thence
$\lim_{n\to\infty}  P_n^D(z;2\lfl an/2\rfl)
=\lim_{n\to\infty}  P_n^T(z;2\lfl an/2\rfl) =0$ if $z\in(0,2]$.

\subsection{Approximating the pressure}

The partition functions $T_n(z)$ and $D_n(z)$ will be
approximated by using the Stirling approximation for the factorial
\begin{equation}
n! = \sqrt{2\pi}\, n^{n+1/2} e^{-n} \L 1 + \sfrac{1}{12n} 
+ O\L \sfrac{1}{n^2}\R \R 
\label{eqn13}  
\end{equation}
to approximate the binomial coefficients.  The summations will be
approximated by an integral which itself will be estimated using a
saddle point method.

\begin{figure}[t!]
\centering\hfil
\input{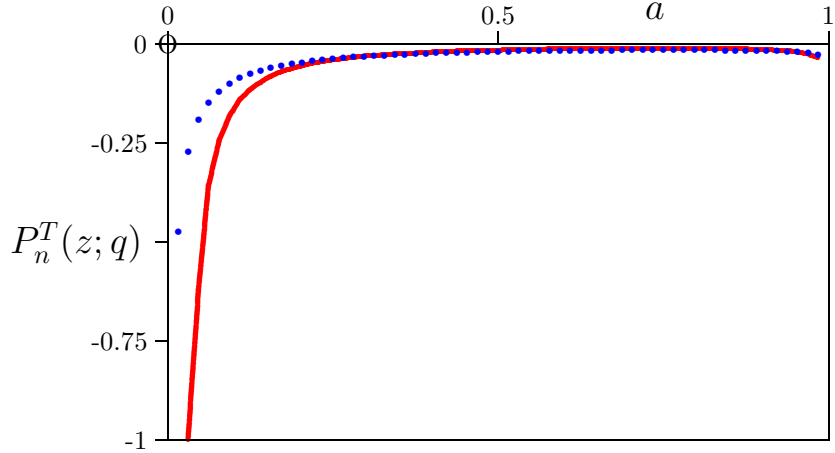}
\caption{The pressure $P_n^T(z;q)$ for directed paths in
$\IntN^2_+$ for $n=128$, $z=3/2$ and $q=2\lfl a\,n/2\rfl$,
plotted as a function of $a$.  The magnitude of the pressure is large close to the origin
and decays with increasing $a$ towards the free endpoint of the
path.  The curve is the asymptotic expression for
$P_n^T(z;q)$ as in equation \Ref{eqnZZ45}.  The dotted curve
is the exact computed pressure, as determined from the partition
functions in equations \Ref{eqn1} and \Ref{eqn2}.}
\label{figureB}  
\end{figure}

Substitute the Stirling approximation in the summands of
equations \Ref{eqn1} and \Ref{eqn2}, take logarithms and simplify
the results. This gives
\begin{equation}
1+\log \L
{\frac { \left( z-1 \right) ^{m}{2}^{n+1}
\sqrt {2}{n}^{n}\sqrt {n} \left( n-2\,m \right) ^{
m-1/2\,n} \left( 2\,m+1 \right) }{\sqrt {
\pi } \left( n+2\,m+2 \right) ^{m+3/2+1/2n}
\sqrt {n-2\,m}}} \R.
\end{equation}
Fixing $n$, there is an $m = \lfl \delta n \rfl$
which maximizes the above (for $\delta \in [0,1/2]$).
The parameter $\delta$ is determined by an asymptotic
expansion in $1/n$:  Put $n=1/\eps$, $m=\delta/\eps$ in the
summand, expand the result in $\eps$ to leading order 
and simplify.  This gives
\begin{equation}
\log \L 
\frac{2}{\eps} 
\L \frac{1-2\delta}{\eps} \R^{(2\delta-1)/2}
\L \frac{1+2\delta}{\eps} \R^{-(2\delta+1)/2}
(z-1)^\delta \R .
\end{equation}
Take the derivative with respect to $\delta$ and solve for $\delta$ to 
find the saddle point at 
\begin{equation}
\delta = \max\LC \frac{z-2}{2z},0 \RC,
\label{eqn9}  
\end{equation}
where we note that $\delta$ cannot be negative.

A similar approach in the case of $T_n(z)$ gives exactly
the same result for the location of the saddle point, as expected.

\subsection{Case 1: $\mathbf{1 < z < 2}$.}
The general approach is to approximate the summations in
equation \Ref{eqn1} and \Ref{eqn2} by integrals
\begin{equation}
T_n(z) \simeq \int_{0}^{\lfl n/2 \rfl} \Bi{n}{\lcl n/2\rcl+m}\,(z-1)^m\,dm
\label{eqn10} 
\end{equation}
for adsorbing directed paths, and
\begin{equation}
D_n(z) \simeq \int_{m=0}^{\lfl n/2 \rfl} \frac{4m+2}{n+2(m+1)}
\Bi{n}{\lfl n/2\rfl+m}\,(z-1)^m\, dm
\label{eqn11} 
\end{equation}
for adsorbing Dyck paths (and where the binomial coefficients are
generalised to continuous values of $m$ by replacing factorials by
Gamma functions). 

\begin{figure}[t!]
\centering\hfil
\input{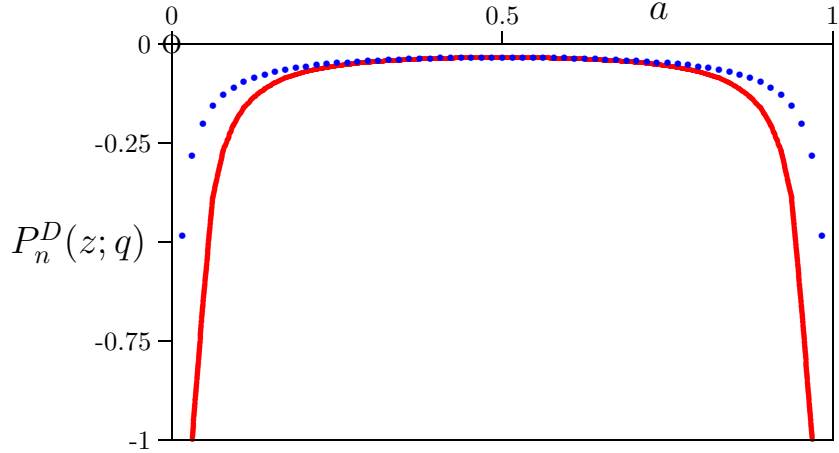}
\caption{The pressure $P_n^D(z;q)$ for Dyck paths 
for $n=128$, $z=3/2$ and $q=2\lfl a\,n/2\rfl$,
plotted as a function of $a$.  The magnitude of the pressure is large close to the 
endpoints of the path, but is smaller for values of $a$ towards the middle region 
of the path.  The curve is the asymptotic expression for
$P_n^D(z;q)$ as in equation \Ref{eqnZZ46}.  The dotted curve
is the exact computed pressure, as determined from the partition
functions in equation \Ref{eqn2}.}
\label{figureC}  
\end{figure}

For asymptotic values of $n$, the integrals above are dominated
by values of $m$ close to the saddle points which are located
close to $m=0$ by equation \Ref{eqn9} and which have a 
spread of $\sqrt{n}$ above $m=0$.

The approximation of $T_n(z)$ and $D_n(z)$ are achieved by 
substituting $n = 1/\eps^2$ and $m=\alpha/\eps = \alpha \sqrt{n}$ 
in the summands above.  Expand the results in $\eps$, and simplify.

In the case of $T_n(z)$, the above prescription results in
\begin{equation}
T_n (z) \simeq \sfrac{2^{3+1/\eps^2}\eps}{\sqrt{2\pi} }
\int_0^{\infty}(z-1)^{\alpha/\eps}\,e^{-2\alpha^2}\,\alpha\,d\alpha .
\end{equation} 
Evaluating the integral gives
\begin{equation}
T_n (z) \simeq 2^{n-1} e^{(n \log^2 (z-1))/8}
\L 1+ \hbox{erf}\L\sqrt{2n}\, \LV \log (z-1) \RV / 4 \R \R .
\end{equation}
An asymptotic expansion of this in $n$ and keeping only 
leading order terms gives
\begin{equation}
T_n (z) \simeq \frac{2^{n+1}}{\sqrt{\pi n}\, \LV \log (z-1) \RV}.
\end{equation}

The same procedure may be applied to $D_n (z)$:  This shows that
\begin{equation}
D_n (z) \simeq \sfrac{2^{1+1/\eps^2}}{\sqrt{2\pi} }
\int_0^{\infty}(z-1)^{\alpha/\eps}\,e^{-2\alpha^2}\,d\alpha .
\end{equation} 
Evaluating the integral produces
\[
\hspace{-2.5cm}
D_n (z) \simeq 2^{n+1/2}
\L \sfrac{1}{\sqrt{\pi n}} +  \LV \log (z-1) \RV
e^{n \log^2 (z-1) / 8} \L 1
+\hbox{erf} \L \sqrt{2n} \LV \log (z-1) \RV /4 \R \R \R . 
\]
Expanding this asymptotic in $n$ and simplifying gives
\begin{equation}
D_n (z) \simeq \frac{2^{n+3}}{\sqrt{2\pi n^3} \,
\log^2 (z-1) } .
\end{equation}

The above results produces the following approximations for the pressures: 
\begin{equation}
\fl
 P_n^T(z;2\lfl an/2\rfl)  \simeq \log \L
1 - \frac{8}{\sqrt{2\pi n^3 a^3 (1-a)}\, \log^2 (z-1) } \R
\label{eqnZZ45}   
\end{equation}
for directed paths, and
\begin{equation}
\fl
 P_n^D(z;2\lfl an/2\rfl) \simeq \log \L
1 - \frac{8}{\sqrt{2\pi n^3 a^3 (1-a)^3}\, \log^2 (z-1) } \R
\label{eqnZZ46}   
\end{equation}
for Dyck paths.  For large values of $n$ the logarithms may be expanded,
and the results should be compared to equations \Ref{eqnDD1} and
\Ref{eqnDD2} for $z=1$.

In figure \ref{figureB} the approximation for 
$ P_n^T(z;2\lfl an/2\rfl)$ (solid curve) is compared with the 
exact values (dotted curve) for $n=128$ and $z=3/2$.  Observe that 
as $n\to\infty$, then $ P_n^T(z;2\lfl an/2\rfl)$ approaches zero.

Expanding the expression for the pressures asymptotically in $n$
gives the expressions for $z<2$ in equations \Ref{eqn20AA} and
\Ref{eqn21AA}.

\subsection{Case 2: $\mathbf{z=2}$.}

\begin{figure}[t!]
\centering\hfil
\input{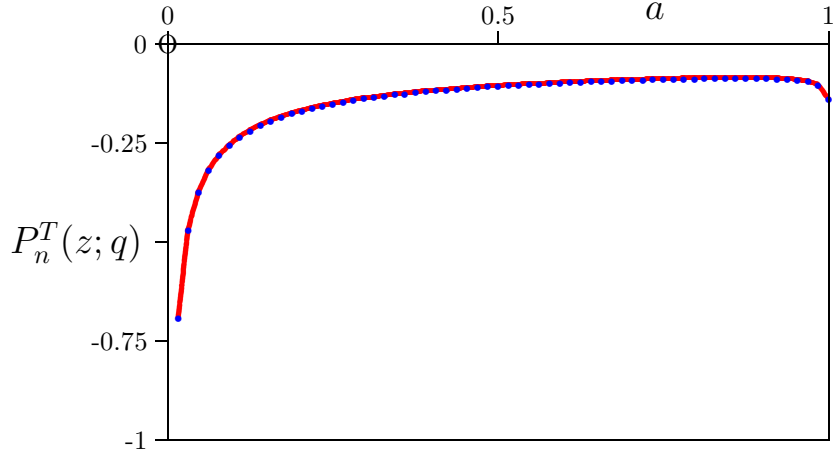}
\caption{The pressure $P_n^T(z;q)$ for directed paths in
$\IntN^2_+$ for $n=128$, $z=2$ and $q=2\lfl a\,n/2\rfl$,
plotted as a function of $a$.  The pressure is large close to the origin
and decays with increasing $a$ towards the free endpoint of the
path.  The curve and dotted points are determined from the
exact expressions for the pressure, which is obtained from 
equations \Ref{eqn1} and \Ref{eqn2} on the one hand,
and equation \Ref{eqnZZ43} on the other hand.}
\label{figureD}  
\end{figure}

The partition functions evaluate exactly in terms of factorials
and Gamma functions if $z=2$:
\begin{equation}
\fl
T_n(2) = 2^{n-1} 
\L \frac{\Gamma\L \sfrac{n+1}{2}\R}{
\sqrt{\pi}\,\Gamma\L \sfrac{n+2}{2}\R} + 1\R,
\q\hbox{and}\q D_n(2) = \frac{n!}{(n/2)! (n/2)!}.
\label{eqnZZ43}  
\end{equation}
Substituting these expressions into equations \Ref{eqn5}
and \Ref{eqn6} gives exact expressions for the pressure, which
we do not reproduce here.  Observe that as $n\to \infty$,
then it follows from the above that $\lim_{n\to\infty}
P_n^D(2;2\lfl an/2 \rfl) = \lim_{n\to\infty} P_n^T(2;2\lfl an/2 \rfl) = 0$. 
The exact presssure $P^T_{n} (2; 2\lfl an/2 \rfl)$ is plotted for
$n=128$ in figure \Ref{figureD}. 

Asymptotic expressions for the pressure can be determined by 
using Stirling's approximation in equations \Ref{eqnZZ43}.  Substitution
and simplification gives 
\begin{eqnarray}
\fl
P_n^T(2;2\lfl a\,n/2\rfl) 
&= \log \L 1 - \frac{2}{\sqrt{2\pi n a}}
    + \frac{2(1-a-\sqrt{1-a})}{\pi n \sqrt{a}\,(1-a)} + O\L n^{-3/2} \R \R, & 
\label{eqnDD50}  
\\
\fl
P_n^D(2;2\lfl a\,n/2 \rfl)
&= \log \L 1-\frac{2}{\sqrt{2\pi na(1-a)}}  + O\L n^{-3/2} \R \R &
\label{eqnDD51}  
\end{eqnarray}
for the pressure in these models.  Expanding the pressures asymptotically
in $n$ gives the results for $z=2$ in equations \Ref{eqn20AA} and
\Ref{eqn21AA}.

\subsection{Case 3: $\mathbf{z>2}$.}

Asymptotic expressions in this regime are obtained by exploring the
saddle point in the summands of $D_n(z)$ and $T_n(z)$.  This 
saddle point is located at $\delta n$ in the asymptotic regime 
(where $\delta$ is given by $(z-2)/2z$ in equation \Ref{eqn9}).

The width of the saddle is proportional to $\sqrt{n}$.  Hence,
use the Stirling approximation in the summands, put
$m = \L\sfrac{z-2}{2z}\R n + \alpha\sqrt{n}$ and
$n = 1/\eps^2$, expand in $\eps$ to $O(1)$, and simplify.
This gives the saddle point approximations
\begin{equation}
T_n(z) \simeq 
\frac{\eps\sqrt{n}}{\sqrt{2\pi}} 
\L \frac{z}{\sqrt{z-1}} \R^{\frac{\eps^2+1}{\eps^2}}
\int_{-\infty}^{\infty} e^{-\frac{\alpha^2 z^2}{2(z-1)}} \, d\alpha 
\end{equation}
and
\begin{equation}
D_n(z) \simeq 
\frac{\eps\sqrt{n}}{\sqrt{2\pi}} 
\L \frac{z-2}{z-1} \R
\L \frac{z}{\sqrt{z-1}} \R^{\frac{\eps^2+1}{\eps^2}}
\int_{-\infty}^{\infty} e^{-\frac{\alpha^2 z^2}{2(z-1)}} \, d\alpha .
\end{equation}
Integrating $\alpha$ over $(-\infty,\infty)$ gives the saddle point
approximations
\begin{equation}
T_n(z) \simeq \L \frac{z}{\sqrt{z-1}}\R^n,\q\hbox{and}\q
D_n(z) \simeq \frac{z-2}{z-1} \L \frac{z}{\sqrt{z-1}}\R^n .
\end{equation}
These results are not unexpected, since substitution and
simplification into equations \Ref{eqn5} and \Ref{eqn6} give
the results in lemmas \ref{lemma1} and \ref{lemma2}.

\begin{figure}[t!]
\centering\hfil
\input{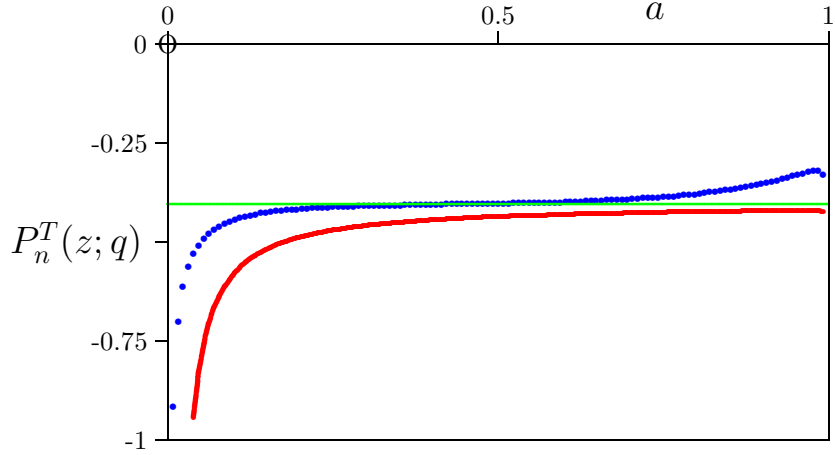}
\caption{The pressure $P_n^T(z;q)$ for directed paths in
$\IntN^2_+$ for $n=256$, $z=5/2$ and $q=2\lfl a\,n/2\rfl$,
plotted as a function of $a$.  The pressure is large close to the origin
and decays with increasing $a$ towards the free endpoint of the
path.  The curve is determined using the asymptotic expressions
in equations \Ref{eqnZZ49} and \Ref{eqnZZ50}.  The dotted 
curve is the exact pressure (determined from equations \Ref{eqn1}
and \Ref{eqn2}).}
\label{figureE}  
\end{figure}

This also shows that the expansion to $O(1)$ in $\eps$ does
not produce expressions which give corrections for
finite $n$ effects.

Finite $n$ corrections to the pressures are obtained by improving
the saddle point approximations above.  Expanding to $O(\eps)$
instead gives the approximations
\begin{equation}
T_n(z) \simeq 
\frac{\eps\sqrt{n}}{\sqrt{2\pi}} 
\L \frac{z}{\sqrt{z-1}} \R^{\frac{\eps^2+1}{\eps^2}}
\int_{-\infty}^{\infty} 
e^{-\frac{\alpha^2 z^2}{2(z-1)} - A_T \eps} \, d\alpha 
\end{equation}
where
\[ A_T = 
\frac{\L \alpha^2 z^4 -(4\alpha^2-3)z^3 + (4\alpha^2+9)z^2 - 18\,z +12
 \R \alpha z}{6(z-1)^2(z-2)}  \]
and
\begin{equation}
D_n(z) \simeq 
\frac{\eps\sqrt{n}}{\sqrt{2\pi}} 
\L \frac{z-2}{z-1} \R
\L \frac{z}{\sqrt{z-1}} \R^{\frac{\eps^2+1}{\eps^2}}
\int_{-\infty}^{\infty} 
e^{-\frac{\alpha^2 z^2}{2(z-1)} - A_D \eps} \, d\alpha 
\end{equation}
where
\[ A_D = 
\frac{\L \alpha^2 z^3 - (2\alpha^2 + 3)z^2 + 9\,z - 6 \R
\alpha z}{6(z-1)^2}   .\]
Expanding $e^{-A_T \eps} = 1 - A_T\eps + \sfrac{1}{2} A_T^2\eps^2
+ O(\eps^3)$ and $e^{-A_D \eps} = 1 - A_D\eps + \sfrac{1}{2} A_D^2\eps^2
+ O(\eps^3)$ in the above and integrating to $\alpha$ should
give improved asymptotic expressions for $T_n(z)$ and $D_n(z)$
if $z$ is not too large.  Simplification gives
\begin{equation}
T_n(z) \simeq 
\frac{z^2+4(3n-1)z-4(3n-1)
}{12n(z-1)}\L\frac{z}{\sqrt{z-1}} \R^n .
\label{eqnZZ49}  
\end{equation}
and 
\begin{equation}
\hspace{-2.5cm}
D_n(z) \simeq 
\frac{z^4+4(3n-2)z^3-30(2n-1)z^2+16(2z-1)(3n-1)
}{12n(z-1)^2(z-2)}\L\frac{z}{\sqrt{z-1}} \R^n .
\label{eqnZZ50}  
\end{equation}
Substitution of the above in equations \Ref{eqn4} and \Ref{eqn5}
and taking $n\to\infty$ again gives $-\log(z-1)$, the expected
limiting pressure.

\begin{figure}[t!]
\centering\hfil
\input{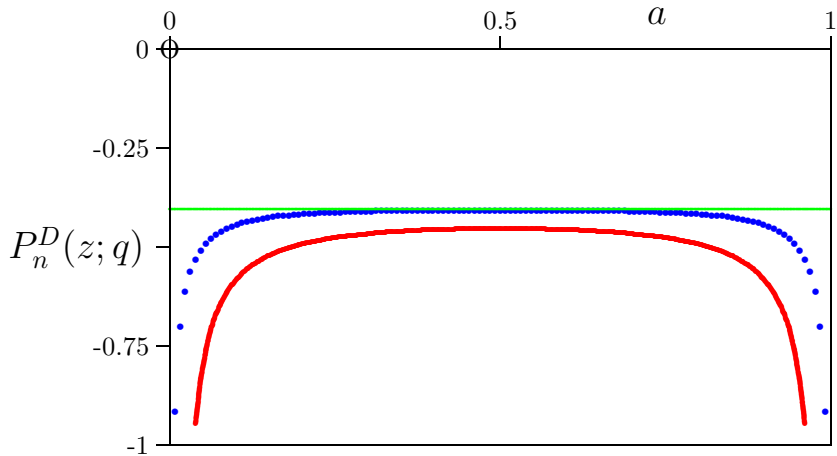}
\caption{The pressure $P_n^D(z;q)$ for Dyck paths in
$\IntN^2_+$ for $n=256$, $z=5/2$ and $q=2\lfl a\,n/2\rfl$,
plotted as a function of $a$.  The pressure is large close to the endpoints
of the path, and smaller towards the center.  The curve is the
asymptotic approximation of the pressure, obtained by using
equation \Ref{eqnZZ50}.  The dotted curve is the exact pressure
(determined from equation \Ref{eqn1} and \Ref{eqn2}).}
\label{figureF}  
\end{figure}

For finite values of $n$ these results show a correction.
In figure \ref{figureE} a plot of $P_n^T(z,2\lfl an/2\rfl)$ 
against $a$ is shown for $n=256$ and $z=5/2$.  Similarly, in
figure \ref{figureF} a plot of the Dyck path case is shown
by plotting $P_n^D(z,2\lfl an/2\rfl)$ as a function of $a$ for $n=256$
and for $z=5/2$.

\section{The forces exerted by adsorbing staircase polygons}

\begin{figure}[t!]
\centering \hfil
\input{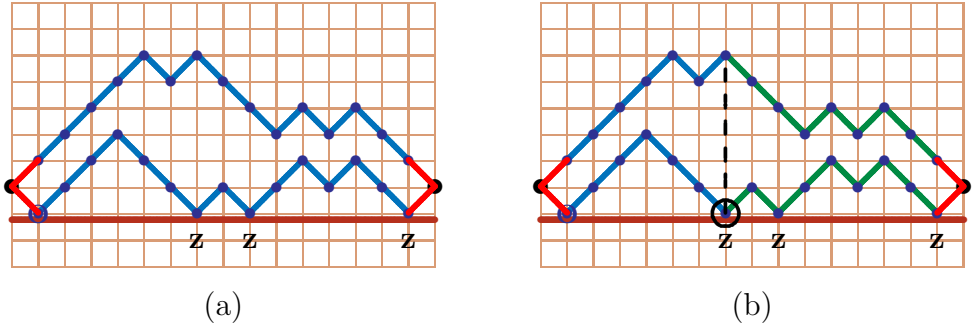}
\caption{A model of staircase polygons.  (a) A staircase polygon with its left-most
and right-most endpoints grafted to the adsorbing line.  If the first
two and last two edges are deleted, then a pair of osculating directed
paths are obtained.  These paths avoid one another, and the bottom
path interacts with the adsorbing line via the activity $z$.  (b)  A staircase
polygon can be cut in a vertical line through one of the visits of the bottom path 
to the adsorbing line into a left and a right pair of paths.  If the first two
and last two edges of the polygon is removed, then each pair of paths 
consists of a Dyck path below a directed path.}
\label{figureG}  
\end{figure}

A staircase polygon adsorbing in the upper half plane is
illustrated in figure \ref{figureG}(a).  If the left-most
and right-most vertex in the staircase polygon is deleted, then
it becomes a pair of directed paths in the upper half plane,
one path below the other, and avoiding each other.  

The bottom path is assumed to start in the origin, and by geometry
the top path starts in the vertex with coordinates $(0,2)$.
In this situation we say that the staircase polygon is attached
to the $X$-axis.  Observe that if the final vertex in the
bottom path has coordinates $(X,Y)$, then the final vertex
in the top path necessarily has coordinates $(X,Y+2)$.

The bottom path may visit the $X$-axis, and these \textit{visits}
are weighted by $z$.  By convention, the visit at the origin 
is not weighted.

In this model we assume that the bottom path always ends in
the $X$-axis in a point with coordinates $(2n,0)$ -- these are
\textit{grafted} staircase polygons.  

Since we have reduced the 
model to two directed paths, we relax the conditions above and 
assume that the top path ends in a point with coordinates $(2n,2j+2)$.
In this case the partition function of the model is known \cite{BEO98}.
This generalisation will be useful, since it will enable use to
compute the partition function of staircase polygons passing
through a point $x$ in the adsorbing line (or hard wall).

Hence, consider two directed paths in the square lattice, 
avoiding one another, and with steps $(1,1)$ and $(1,-1)$.
Suppose the first path is a Dyck path starting at the origin
and terminating in the vertex with coordinates $(2n,0)$.
Suppose furthermore that visits of the first path to the
$X$-axis are weighted by $z$.  

\begin{figure}[t!]
\centering \hfil
\input{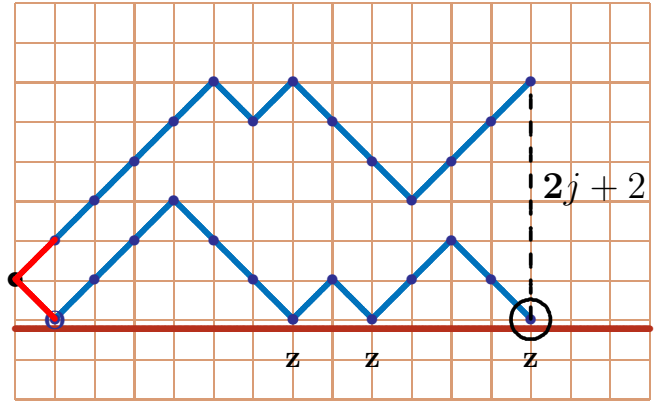}
\caption{A Dyck path below a direct path.  By deleting the first two
edges of this pair of paths, two directed paths which avoids one another
are obtained.  The bottom path is constrained to end in the adsorbing line;
it is a Dyck path, and it adsorbs via the activity $z$ in the adsorbing line.
The top path is a directed path, and its endpoint has height $2j+2$ above
the adsorbing line.}
\label{figureH}  
\end{figure}

Let the second path start
in the vertex with coordinates $(0,2)$ and terminate
in the vertex with coordinates $(2n,2j+2)$, as illustrated
in figure \ref{figureH}.  

Then the partition function of the pair
of paths is
\begin{eqnarray}
\fl
C_n(z;j) = \sum_{\m1=0}^n \sum_{\m2=0}^j
K(\m1,\m2,j,n) \times & & \label{eqn33} 
\\
 \qq\qq\Bi{2n+3}{n+\m2+2} \Bi{2n+3}{n+\m1+j+3} (z-1)^{\m1+\m2} & &
\nonumber
\end{eqnarray}
where
\[ \fl
K(\m1,\m2,j,n)
= \frac{(2\m2+1)(2\m1+2j+3)(\m1+\m2+j+2)(\m1-\m2+j+1)
}{(2n+1)(2n+2)(2n+3)^2} .\]
One may check that $C_0(z;0)=1$, $C_1(z;0)=z$, $C_1(z;1) = z$ 
and $C_2(z;0)=z+2z^2$. For a derivation of this result, see for 
example references \cite{BEO98,JvR99,JvR00} but note the 
misprint in those expressions when compared to the above. 
Observe that the length of each path is $2n$ so that $n$
is the half-length of each path.  Similarly, the distance
between the endpoints is $2j+2$, so that $j+1$ is the 
half-distance between the endpoints.


The pressure on a vertex with coordinates $(q,0)$ in the $X$-axis can 
be computed by determining the partition function of pairs
of paths which avoids this vertex.  This, in turn, can be done
if one first determines the partition function of pairs of paths
which passes through the vertex with coordinates $(q,0)$.  This\
situation is illustrated in figure \ref{figureG}(b), where the bottom path
is constrained to pass through the marked visit.  By cutting the
polygon into two parts in the vertical line which passes through the
visit, the pair of paths are divided into two sets of two paths
each (one pair stepping from the left, and the other from the right), 
and each pair with endpoints a distance $2j+2$ apart as illustrated 
in figure \ref{figureH}. (Reflect the paths on the right hand
side of the cut to get it in the same orientation).

In other words, the partition function
of the model in figure \ref{figureG}(b) is given by
\begin{equation}
Z^*_n (z;q) = \sum_{j=0}^n C_{q} (z;j) \, C_{n-q} (z;j) 
\label{eqn35Q}  
\end{equation}
The partition function of staircase polygons avoiding the 
visit $(q,0)$ is given by
\begin{equation}
Z^\bullet_n(z;q) = C_n(z;0) - Z_n^* (z;q) .
\end{equation}
These give an expression for the net pressure on the vertex
with coordinates $(q,0)$:
\begin{equation}
\fl
P_n^S(z;q) = \log(C_n(z;0)) - \log(Z^\bullet_n(z;q))
= \log \L 1 - \frac{Z_n^*(z;q)}{C_n(z;0)} \R .
\label{eqn37Q}  
\end{equation}
In other words, to determine the force, one must compute
both $C_n(z;0)$ and $Z_n^*(z;q)$. 

In principle, $P_n^S(z;q)$ can be determined exactly from the
results above, but these expressions are complicated and not
very informative.  Hence, we shall approximate them to both
estimate the pressure at finite values of $n$, and in the case that
$n\to\infty$.  As in the case of directed walks, there will be
different results for $z>2$ (the adsorbed phase) and $z\leq 2$.

\subsection{$\mathbf z=2$:}

In the case that $z=2$ and $j=0$, equation \Ref{eqn33} simplifies to
\begin{equation}
C_n(2;0) = \LH \frac{2(2n+1)}{n+2}\RH \C{C}_n^2
\end{equation}
where $\C{C}_n = \sfrac{1}{n+1} \Bi{2n}{n}$ is Catalan's number. 

Summing $j=0$ to $n$ instead in equation \Ref{eqn33} gives
\begin{equation}
\sum_{j=0}^n C_n(z;j) = \frac{\Gamma\L n+\sfrac{1}{2}\R 16^n}{
\sqrt{\pi}\, \Gamma(n+2)}.
\end{equation}

More generally, for fixed $j$, $C_n(2;j)$ simplifies to
\begin{equation}
C_n(2;j) = \LH\frac{2(2n+1)(j+1)^2\C{C}_n}{(n+j+1)(n+j+2)}\RH \Bi{2n}{n+j} .
\end{equation}
These results can be used to obtain exact expressions for 
the force $P_n^S(2;q)$ in equation \Ref{eqn37Q}.

In particular, one may evaluate $Z_n^*(2;q)$ in equation \Ref{eqn35Q}
exactly to
\begin{equation}
\fl
Z_n^* (2;q) = \frac{
2(n(3q+2)-3q^2+1)\Gamma(n+\half)\Gamma(q+\half)\Gamma(n-q+\half)\, 16^n}{
\sqrt{\pi^3} \, \Gamma(q+2)\Gamma(n+3)\Gamma(n-q+2)}.
\label{eqn41}  
\end{equation}
Dividing by $C_n(2;0)$ and computing $P_n^S(2;q)$ in equation
\Ref{eqn37Q} gives
\begin{equation}
\fl
P_n^S(2;q)  = \log \L 1-
\frac{ \L n(3q+2)-3q^2+1 \R \Gamma(q+\half) \Gamma (n-q+\half) \Gamma(n+2) }{
2\sqrt{\pi} \,\Gamma(q+2)\Gamma(n-q+2)\Gamma(n+\frac{3}{2}) } \R .
\label{eqn42}  
\end{equation}
If one put $q=2\lfl an/2 \rfl$ wheren $0<a<1$ and take the limit $n\to\infty$,
then a symbolic computations program \cite{Maple} shows that
$\lim_{n\to\infty} P_n^S(2;2\lfl an/2 \rfl )= 0$. This shows that the
pressure is zero in the limit as $n\to\infty$ and $z=2$.

\begin{figure}[t!]
\centering \hfil
\input{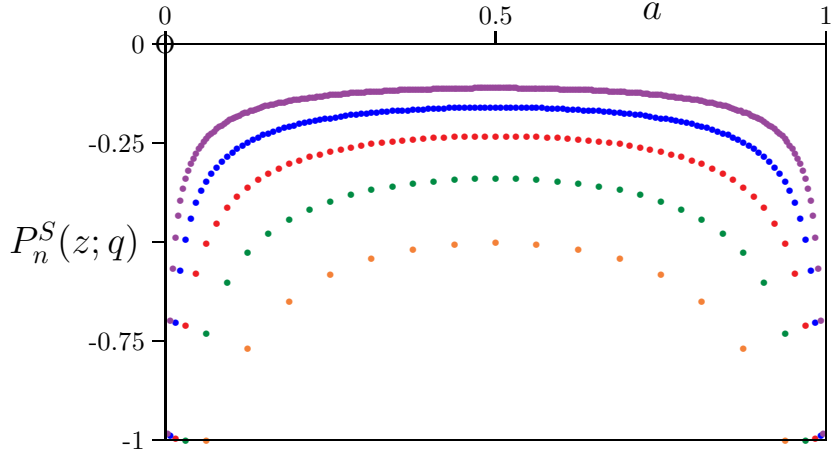}
\caption{Exact pressures for adsorbing staircase polygons with 
$z=2$, $q=2\lfl a\,n/2\rfl$ and for $n\in\{16,32,64,128,256\}$.  The pressure 
is zero in the limit that $n\to\infty$.  These points were computed
by using equation \Ref{eqn42}, which is the exact expression for 
$P_n^S(2;q)$.  }
\label{figureI}  
\end{figure}

\subsection{Asymptotics}

By putting $\m1+\m2 =k$ and summing over $\m1$, equation
\Ref{eqn33} can be cast in the form
\begin{eqnarray}
\fl
C_n(z;j) = \sum_{k=0}^{n} \frac{(j+k+2)}{(n+1)^2} 
 \Bi{2n}{n} \Bi{2n+2}{n+j+k+3} (z-1)^k  & & 
\label{eqn38}  
\\
 \fl \q -\sum_{k=0}^{n} 
  \L \frac{(j+k+2)(n-2j(k+1)-2k-1)}{2(n+1)^2(2n+1)} \R
 \Bi{2n+2}{n+k+2} \Bi{2n+2}{n+j+2} (z-1)^k .& & \nonumber
\end{eqnarray}
This form presents a single summand over $k$ which one may
consider for approximation.

Isolate the summands of $C_n(z;j)$ above and cast them in terms
of Gamma functions.  This gives
\begin{eqnarray}
\fl
C_1 = \frac{8(j+k+2) \Gamma^2\L n+\frac{3}{2}\R (z-1)^k 16^n }{
\pi (n+1)(2n+1)\Gamma\L n+j+k+4\R \Gamma\L n-j-k\R} ,& & 
\label{eqn44}  
\\
\fl
C_2 = \frac{2(j+k+2)\L 2j(k+1) - n +2k+1 \R \Gamma^2 \L 2n+2\R (z-1)^k}{
(2n+1) \Gamma \L n+j+3\R \Gamma \L n-j+1\R 
\Gamma \L n+k+3\R \Gamma \L n-k+1\R }  ,& &
\label{eqn45}  
\end{eqnarray}
where $C_1$ is the summand of the first summation in
equation \Ref{eqn38} and $C_2$ is the summand of the second
summation.

Numerical evaluations of $C_1$ and $C_2$ for given $n$ indicate that
the summands are large for small $k \ll j$ (and $j$ is also 
small) for $z<2$, at both $k$ and $j$ small for $z=2$ and at $k\gg j$ 
for $z>2$ (and both $k$ and $j$ are large). In addition, both sums 
in the above make a substantial contribution
of opposite signs, and so both must be examined to determine suitable
asymptotics.

Proceed by determining the dominant terms in the summands
of equation \Ref{eqn38}.  The binomial coefficients will be
approximated by using the Stirling approximation for factorials
(see equation \Ref{eqn13}).  Take logarithms of the first
summand, substituting $n=1/\eps$, $j=\delta/\eps$ and $k=\alpha/\eps$
and expand the resulting expression in $\eps$.  The leading term is
\begin{equation}
\log \L \frac{16(1-\delta-\alpha)^{\delta+\alpha-1}}{
(\delta+\alpha+1)^{\delta+\alpha+1}}\,  (z-1)^\alpha \R .
\end{equation}
Taking the derivative and solving for $\alpha$ gives
\begin{equation}
\alpha_m = \max\{\sfrac{z-2}{z} - \delta,0\}
\label{eqn48} 
\end{equation}
since $\alpha_m$ cannot be negative.

A similar treatment of the summand in the second sum in
equation \Ref{eqn38} gives the leading term in an expansion in $\eps$:
\begin{equation}
\log \L  \frac{16(1-\delta)^{\delta-1}(1-\alpha)^{\alpha-1}}{
(1+\alpha)^{\alpha+1} (1+\delta)^{\delta+1} } \, (z-1)^\alpha .\R
\end{equation}
Taking the derivative and solving for $\alpha$ gives
\begin{equation}
\alpha_M = \max\{\sfrac{z-2}{z},0\}
\label{eqn50} 
\end{equation}
since $\alpha_m$ cannot be negative.

\subsubsection{$z<2$:}
Proceed by approximating the summands $C_1$ and $C_2$ above.  Take 
logarithms of $C_1$ and $C_2$, substitute the Stirling approximation
for factorials (see equation \Ref{eqn13}) and simplify the results.
Since the summands are dominated by $k\ll j$, and $j$ small, substitute
$j=\delta/\eps$ and $n=1/\eps^2$ and expand to $O(\eps)$. Summing
over $k$ and replacing $\eps = 1/\sqrt{n}$ and $\delta = j/\sqrt{n}$
show that to leading order $C_1$ and $C_2$ cancel.

This shows that higher order terms must be determined in this case.
Expanding to $O(\eps^{12})$, summing over $k$ and combining 
the contributions of $C_1$ and $C_2$ and then extracting the
leading order terms give
\begin{equation}
\fl
\sum_k \L C_1 + C_2 \R \simeq
\frac{4(j+1)\L 2j^2(2-z) + j(8-z) + 6\R e^{-j^2/n} 
\,z\, 16^n}{\pi n^5 (z-2)^4} .
\end{equation}
One may extract the asymptotic behaviour for $j=0$ and $j= O(\sqrt{n})$.
The above can be simplified taking only the
fastest growing terms in each factor.  This shows that
\begin{equation}
\fl
C_n(z;j) = \sum_k \L C_1 + C_2 \R \simeq \cases{
\frac{24\,z\,16^n}{\pi n^5 (2-z)^4}, & \hbox{if $j=0$;} \\
\frac{8\,z\,j^3\,e^{-j^2/n}\,16^n}{\pi n^5 (2-z)^3}, 
& \hbox{if $j= O(\sqrt{n})$.}   }
\end{equation}
The partition function in equation \Ref{eqn35Q} can be approximated
from these last expressions, and similarly, one may approximate
$C_n(z;0)$ in equation \Ref{eqn33}.  This finally gives an approximation
for $P_n^S (z;q)$ in equation \Ref{eqn37Q}. 

In particular, one obtains that
\begin{equation}
\fl
\sum_{j=0}^n C_q(z;j) C_{n-q}(z;j)
\simeq \int_0^\infty 
\L \frac{8\,z\,j^3\,e^{-j^2/q}\,16^q}{\pi q^5 (2-z)^3} \cdot
\frac{8\,z\,j^3\,e^{-j^2/(n-q)}\,16^{n-q}}{\pi (n-q)^5 (2-z)^3} \R dj .
\end{equation}
The integral can be readily done, and after division by the 
asymptotic expression for $C_n(z;0)$ one is left with the following
approximation for the force:
\begin{equation}
P_n^S(z;q) \simeq \log \L 1-
\frac{5\, z\, n^{3/2}}{2\sqrt{\pi}\,\L q(n-q)\R^{3/2} (2-z)^2 } \R .
\end{equation}
Replacing $q$ by $2\lfl a\,n/2 \rfl$ and simplifying then gives
\begin{equation}
P_n^S(z;2\lfl a\,n/2\rfl) \simeq \log \L 1-
\frac{5\, z}{2n^{3/2}\sqrt{\pi a^3(1-a)^3} \, (2-z)^2} \R .
\label{eqn55}  
\end{equation}
Hence, for $a\in(0,1)$ $P_n^S(z;2\lfl a\,n/2\rfl) \to 0$ as $n\to\infty$.
By expanding the logarithm, the case that $z<2$ in equation
\Ref{eqn22AA} is obtained.

\begin{figure}[t!]
\centering\hfil
\input{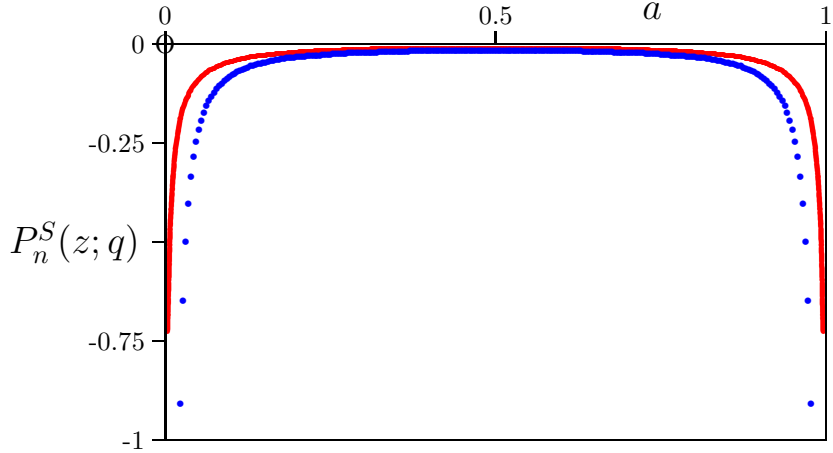}
\caption{The pressure $P_n^S(z;q)$ for staircase polygons in
$\IntN^2_+$ for $n=256$, $z=\frac{3}{2}$ and $q=2\lfl a\,n/2\rfl$,
plotted as a function of $a$.  The pressure is large close to the endpoints
of the polygons.  The curve is the approximate pressure 
(equation \Ref{eqn55}) while the dotted curve are the exact
values of the pressure.}
\label{figureJ}  
\end{figure}

In figure \ref{figureJ} the  pressure $P_n^S(z;2\lfl a\,n/2\rfl)$ is plotted for
$a\in(0,1)$,  $z=3/2$ and $n=256$ comparing the approximate
expression in equation \Ref{eqn55} against the exact calculated
pressure.

\subsubsection{$z=2$:}

\begin{figure}[t!]
\centering\hfil
\input{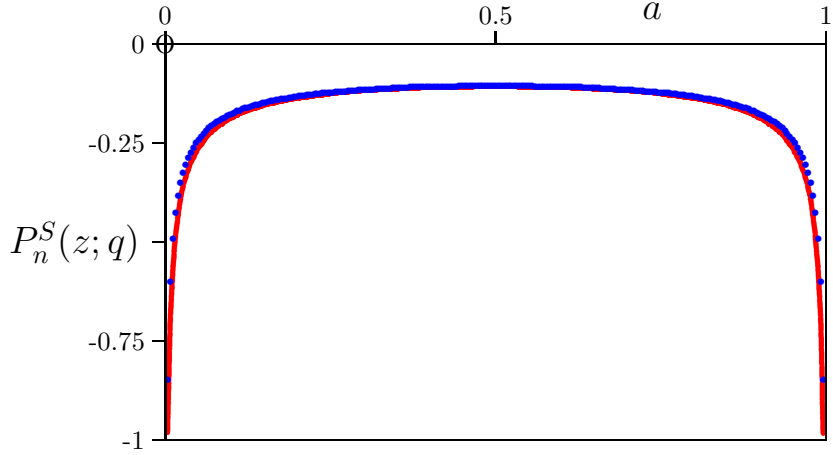}
\caption{The pressure $P_n^S(z;q)$ for staircase polygons in
$\IntN^2_+$ for $n=256$, $z=2$ and $q=2\lfl a\,n/2\rfl$,
plotted as a function of $a$.  The pressure is large close to the endpoints
of the polygons.  The curve was determined from equation
\Ref{eqn42} and the points along the dotted curve ares the exact values computed
from the partition function in equation \Ref{eqn38}.}
\label{figureK}  
\end{figure}

The case $z=2$ has been solved explicitly, as seen in equations
\Ref{eqn41} and \Ref{eqn42}.  The expression for $P_n^S(2;q)$ in
equation \Ref{eqn42} is in terms of  Gamma functions, and these
can be approximated when $q=2\lfl a\,n/2\rfl$ for large $n$ using 
Stirling's approximation.  This shows that
\begin{equation}
P_n^S(2;2\lfl a\,n/2\rfl) = \log \L 1-
\frac{3}{2 \sqrt{n\pi a(1-a)} } + O\L n^{-3/2} \R  \R .
\label{eqn81}   
\end{equation}
By expanding the logarithm, the case that $z=2$ in equation
\Ref{eqn22AA} is obtained.
In figure \ref{figureK} a plot of the approximation and exactly
calculated values of the pressure is presented.

\begin{figure}[b!]
\centering\hfil
\input{figureL.tex}
\caption{The pressure $P_n^S(z;q)$ for staircase polygons in
$\IntN^2_+$ for $n\in\{32,64,128,256\}$, $z=3$ and $q=2\lfl a\,n/2\rfl$,
plotted as a function of $a$.  The pressure is large close to the endpoints
of the polygons.  The solid curve is the exact pressure at $n=256$ 
(determined from equation \Ref{eqn38}) while the set of dotted curves
are the asymptotic results (equation \Ref{eqnZZ83}) for 
$n$ doubling from $32$ to $256$.}
\label{figureL}  
\end{figure}

\begin{figure}[t!]
\centering\hfil
\input{figureM.tex}
\caption{The pressure $P_n^S(z;q)$ for staircase polygons in
$\IntN^2_+$ for $n\in\{32,64,128,256\}$, $z=4$ and $q=2\lfl a\,n/2\rfl$,
plotted as a function of $a$.  The pressure is large close to the endpoints
of the polygons.  The solid curve is the exact pressure at $n=256$ 
(determined from equation \Ref{eqn38}) while the set of dotted curves
are the asymptotic results (equation \Ref{eqnZZ83}) for 
$n$ doubling from $32$ to $256$.}
\label{figureM}  
\end{figure}

\begin{figure}[b!]
\centering\hfil
\input{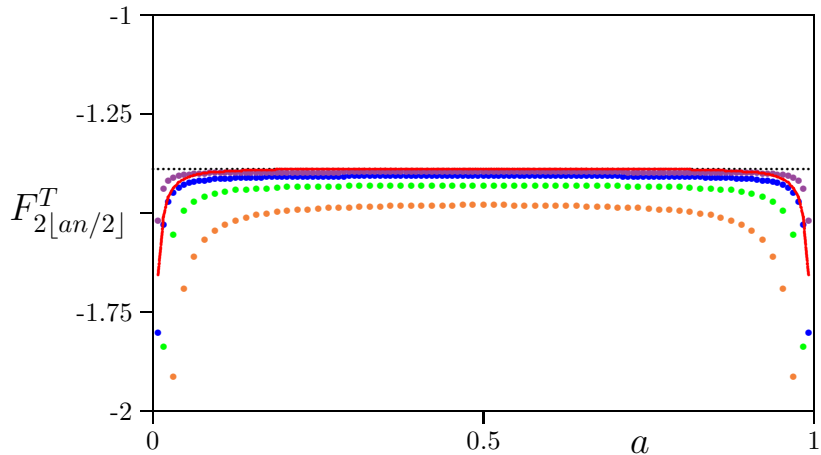}
\caption{The pressure $P_n^S(z;q)$ for staircase polygons in
$\IntN^2_+$ for $n\in\{32,64,128,256\}$, $z=5$ and $q=2\lfl a\,n/2\rfl$,
plotted as a function of $a$.  The pressure is large close to the endpoints
of the polygons.  The solid curve is the exact pressure at $n=256$ 
(determined from equation \Ref{eqn38}) while the set of dotted curves
are the asymptotic results (equation \Ref{eqnZZ83}) for 
$n$ doubling from $32$ to $256$.}
\label{figureN}  
\end{figure}

\subsubsection{$z>2$:}
The summands $C_1$ and $C_2$ in equations \Ref{eqn44} and \Ref{eqn45}
provide the starting point.  Numerical experimentation shows that $C_1$
is dominated by terms with $k=O(n)$ and $j= O\L \sqrt{n} \R$ while
$j+k = O(n)$ with a spread of the peak proportional to $\sqrt{n}$.
This is in particular confirmed by the result in equation 
\Ref{eqn48} which indicates that the dominant values of $j$ and 
$k$ in equation \Ref{eqn44} are at $k = ((z-2)/z)n - j$ and
$j = O\L \sqrt{n} \R$.

Hence, put $n=1/\eps^2$, $k=(z-2)/(z\eps^2) - \delta/\eps + \alpha/\eps$
and $j=\delta/\eps$ in equation \Ref{eqn44}.  Take logarithms
and expand in $\eps$ to $O(1)$.  Exponentiating and integrating 
the resulting expression gives the asymptotic expression
\begin{equation}
C_1(n,j) = 
\frac{(z-2) 4^n z^{2n+1}}{\sqrt{\pi} (z-1)^{n+j+3}n^{3/2} }
\L 1 - \sfrac{9}{8n}\L 1+o(1) \R \R
\label{eqn57}   
\end{equation}
where the substitution $\delta = j/\sqrt{n}$ was made.  Expanding
to higher order in $\eps$ before integrating gives subleading
corrections, and does not alter this leading term correction.

A similar approach should give an approximation to $C_2$.  However,
some care is needed in this case.  Determining only leading
term behaviour leads to incorrect results, and it is necessary
to include higher order corrections.  The  arguments leading
to equation \Ref{eqn50} show that the appropriate choices for
$j$ and $k$ is $n=1/\eps^2$ and $k=(z-2)/(z\eps^2) + \alpha/\eps$
and $j=\delta/\eps$.  Substitute this into equation \Ref{eqn45},
and as before, take the logarithm, expand to $O(\eps)$ (rather
than just $O(1)$). Exponentiate the result and expand in $\eps$
to $O(\eps^2)$.   Integrating $\alpha\in(-\infty,\infty)$ for
small $\eps$ then gives the asymptotic expression
\begin{eqnarray}
& \hspace{-1cm} C_2 (n,j) = 
\frac{z-2}{(z-1)^2} \L \frac{z^2}{z-1} \R^n 
\frac{4^n e^{-j^2/n}}{\sqrt{\pi n^3}} \times& \nonumber \\
&\hspace{2cm} \L z(2j+1)-4(j+1) - \sfrac{2(z-2)j^2}{n} + o\L 1/n  \R \R&
\label{eqn58}   
\end{eqnarray}
for the summand $C_2$.

The approximation for $C_1$ peaks sharply at $j=0$ while that for
$C_2$ dominates the contribution from $C_1$ when $j>0$ and peaks 
at a value of $j>0$.  At $j=0$ both terms make a contribution.  
Hence, approximate the pressure at the point $(q,0)$ by
\begin{equation}
\fl
\log \L 1-
\int_{-\infty}^\infty \LH \frac{\L C_1(q,j)+C_2(q,j)\R 
\L C_1(n-q,j)+C_2(n-q,j)\R}{C_1(n,0)+C_2(n,0)} 
\RH \, dj \R .
\label{eqn59}   
\end{equation}
Put $q = 2\lfl a\,n/2\rfl$ and expand the integrand above and integrate term
by term.  Expand the results asymptotically in $n$ and keep terms to 
$O(n^{-2})$.  This finally gives the result
\begin{eqnarray}
& \hspace{-2cm}
P_n^S(z;2\lfl a\,n/2\rfl) \simeq  \log \L \frac{1}{z-1} 
- \frac{3}{8} \frac{4z^2-29z+64}{n(z-1)(z-2)} \L 1+\frac{9z}{8n(z-2)^2} \R
\right.&
\label{eqnZZ83}  
 \\ 
&\hspace{2cm} \left.
 - \frac{3z^2}{2\sqrt{\pi n^3 a^3(1-a)^3} (z-2)(z-1)^3\log(z-1)} \R&.
\nonumber
\end{eqnarray}
Taking $n\to\infty$ shows that $P_n^S(z;2\lfl a\,n/2\rfl)  \to - \log (z-1)$, 
consistent with the result for adsorbing Dyck paths in the $z>2$
regime (see lemma \ref{lemma2}). By factoring $(z-1)$ from the
argument of the logarithm in equation \Ref{eqnZZ83} and expanding
the logarithm asymptotically, the case $z>2$ in equation
\Ref{eqn22AA} is obtained.

\section{Conclusions}

In this paper we have investigated the pressure exerted by a directed
path in a half-space on the $X$-axes.  This is a directed model of the
forces exerted by a two dimensional polymer grafted to a hard wall.
Our first model is of a directed path, in which we considered two cases,
namely an adsorbing Dyck path model with both endpoints grafted to 
the adsorbing line, and a directed path with only one endpoint grafted to
the adsorbing line.

The pressure curve for the Dyck path is both a function of the length
of the path ($n$), and the adsorption activity ($z$).   Asymptotic
expressions were obtained  for the pressure at $(q,0) = (2\lfl an/2 \rfl,0)$ for
$a\in(0,1)$.  The pressure curve is symmetric about $a=1/2$ and 
is asymptotically given by equation \Ref{eqnZZ46} for $z<2$,
equation \Ref{eqnDD51} for $z=2$ and equation \Ref{eqnZZ50}
for $z>2$.  The pressure is zero as $n\to\infty$ and $z\leq 2$, but
there is a net constant limiting pressure of magnitude $\log(z-1)$
for $z>2$. In this regime the adsorbed paths are close to
the hard wall, and the result is a non-vanishing pressure.

Similar observations can be made for the pressure due to an adsorbing
directed path.  In this model, asymptotic expressions for the
pressure were determined and is given in equation \Ref{eqnZZ46}
for $z<2$, and by substituting equation \Ref{eqnDD50} into equation
\Ref{eqn6} for $z=2$, and 
equation \Ref{eqnZZ49} for $z>2$.  The pressure profiles for
finite length paths are not symmetric in this model, but the limiting
pressure is equal to the limiting pressure of Dyck paths. 

We have also determined the pressure due to adsorbing staircase
polygons which were grafted to the adsorbing line at both ends.
Determining the limiting pressure involved more careful analysis,
but the results are given by expressions which are similar to the
Dyck path results.  The pressures are given asymptotically by equation
\Ref{eqn55} for $z<2$,  equation \Ref{eqn81} for $z=2$ and 
equation \Ref{eqnZZ83} for $z>2$.

\vspace{1cm}
\section*{Acknowledgements}
EJJvR acknowledges support in the form of a NSERC Discovery Grant from
the Government of Canada.

\section*{Bibliography}
\bibliographystyle{amsplain}
\bibliography{thomas}

\providecommand{\bysame}{\leavevmode\hbox to3em{\hrulefill}\thinspace}
\providecommand{\MR}{\relax\ifhmode\unskip\space\fi MR }
\providecommand{\MRhref}[2]{%
  \href{http://www.ams.org/mathscinet-getitem?mr=#1}{#2}
}
\providecommand{\href}[2]{#2}
\begin{thebibliography}{10}

\bibitem{BHGMLSW95}
H.D. Bijsterbosch, V.O. de~Haan, A.W. de~Graaf, M.~Mellema, F.A.M. Leermakers,
  M.A. Cohen~Stuart, and A.A. van Well, \emph{Tethered adsorbing chains:
  Neutron reflectivity and surface pressure of spread diblock copolymer
  monolayers}, Langmuir \textbf{11} (1995), 4467--4473.

\bibitem{BEO98}
R.~Brak, J.W. Essam, and A.L. Owczarek, \emph{New results for directed vesicles
  and chains near an attractive wall}, Journal of Statistical Physics
  \textbf{93} (1998), 155--192.

\bibitem{CS95}
M.A. Carignano and I.~Szleifer, \emph{On the structure and pressure of tethered
  polymer layers in good solvent}, Macromolecules \textbf{28} (1995),
  3197--3204.

\bibitem{CNC03}
E.P.K. Currie, W.~Norde, and M.A. Cohen~Stuart, \emph{Tethered polymer chains:
  surface chemistry and their impact on colloidal and surface properties},
  Advances in Colloid and Interface Science \textbf{100--102} (2003), 205--265.

\bibitem{Maple}
Waterloo~Maple Inc, \emph{Maple 12},  (2008).

\bibitem{JvR00}
E.J. Janse~van Rensburg, \emph{The statistical mechanics of interacting walks,
  polygons, animals and vesicles}, vol.~18, Oxford University Press, USA, 2000.

\bibitem{DJ12}
I.~Jensen, W.G. Dantas, C.M. Marques, and J.F. Stilck, \emph{Pressure exerted
  by a grafted polymer on the limiting line of a semi-infinite square lattice},
  Journal of Physics A: Mathematical and Theoretical \textbf{To appear}.

\bibitem{S99}
R.P. Stanley, \emph{Enumerative combinatorics, volume 2}, Cambridge Studies in
  Advanced Mathematics, vol.~62, Cambridge University Press, 1999.

\bibitem{JvR99}
B.~van Rensburg, \emph{Adsorbing staircase walks and staircase polygons},
  Annals of Combinatorics \textbf{3} (1999), 451--473.

\bibitem{W98}
S.G. Whittington, \emph{A directed-walk model of copolymer adsorption}, Journal
  of Physics A: Mathematical and General \textbf{31} (1998), 8797--8804.

\end{thebibliography}

\end{document}